\documentclass[namedreferences]{solarphysics}
\usepackage[hyperref,optionalrh,solaromanenum]{spr-sola-addons} % For Solar Physics 
\usepackage{graphicx}                    % For eps figures, newer & more powerfull
\usepackage{color}                       % For color text: \color command
\usepackage{breakurl}                         % For breaking URLs easily trough lines
\usepackage{paralist}
\newcommand{\etal}{{\it et al.}}

\chardef\us=`\_

\begin{document}

\begin{article}

\begin{opening}

\title{70 years of Sunspot Observations at Kanzelh\"ohe Observatory:
Systematic Study of Parameters Affecting the Derivation of the Relative Sunspot Number}

%%%%%%%%%%%%%%%%%%%%%%%%%%%%%%%%%%%%%%%%%%%%%%%%%%%
%% Authors Names
%
\author[addressref={obs},corref,email={werner.poetzi@uni-graz.at}]{\inits{W.}\fnm{Werner}~\lnm{P\"otzi}}
\author[addressref={obs,igam},corref,email={astrid.veronig@uni-graz.at}]{\inits{A.M.}\fnm{Astrid~M.}~\lnm{Veronig}}
\author[addressref={igam},corref,email={manuela.temmer@uni-graz.at}]{\inits{M.}\fnm{Manuela}~\lnm{Temmer}}
\author[addressref={obs},corref,email={dietmar.baumgarnter@uni-graz.at}]{\inits{D.J.}\fnm{Dietmar~J.}~\lnm{Baumgartner}}
\author[addressref={obs},corref,email={heinrich.freislich@uni-graz.at}]{\inits{H.}\fnm{Heinrich}~\lnm{Freislich}}
\author[addressref={obs},corref,email={heinz.strutzmann@uni-graz.at}]{\inits{H.}\fnm{Heinz}~\lnm{Strutzmann}}
\address[id=obs]{Kanzelh\"ohe Observatory for Solar and Environmental Research, University of Graz, Austria}
\address[id=igam]{Institute of Physics/IGAM, University of Graz, Austria}
%%%%%%%%%%%%%%%%%%%%%%%%%%%%%%%%%%%%%%%%%%%%%%%%%%%
%% Runningheads
%
\runningauthor{W. P\"otzi \etal}
\runningtitle{Sunspot Relative Numbers at Kanzelh\"ohe Observatory}

%%%%%%%%%%%%%%%%%%%%%%%%%%%%%%%%%%%%%%%%%%%%%%%%%%%
%%% Abstract 

\begin{abstract}
Kanzelh\"ohe Observatory (KSO) was founded during World War II
by the ``Deutsche Luftwaffe" (The German Airforce)  as one station of a network of 
observatories, which would provide information on solar activity in order to better
assess the actual conditions of the Earth's ionosphere in terms of radio-wave
propagation. Solar observations began in 1943 with photographs 
of the photosphere and drawings of sunspots, plage regions and faculae, as well as 
patrol observations of the solar corona. At the beginning all data were sent
to Freiburg (Germany). After WW\,II international cooperation was established and
the data were sent to Zurich, Paris, Moscow, and Greenwich. Relative sunspot numbers
are derived since 1944. The agreement between relative sunspot numbers derived at
KSO and the new International Sunspot Number (ISN) \citep{SIDC} lies within $\approx\,10\%$.
However, revisiting the historical data, we also find periods with larger deviations.
The reasons for the deviations were twofold: On the one hand a major 
instrumental change took place during which a relocation and modification of the
instrument happened. On the other hand, a period of frequent replacement of personnel 
caused significant deviations, {\it i.e.} stressing the importance of experienced observers.
In the long term, the instrumental improvements led to better image quality. 
Additionally we find a long term trend towards better seeing conditions since the 
year 2000.
\end{abstract}

%%%%%%%%%%%%%%%%%%%%%%%%%%%%%%%%%%%%%%%%%%%%%%%%%%%
%% Keywords
%

\end{opening}
%-------------------------------------------------

%%%%%%%%%%%%%%%%%%%%%%%%%%%%%%%%%%%%%%%%%%%%%%%%%%%
%% Sections
%
\section{Introduction}%\label{s:introduction} 

\begin{figure} 
 \centerline{\includegraphics[width=1\textwidth,clip=]{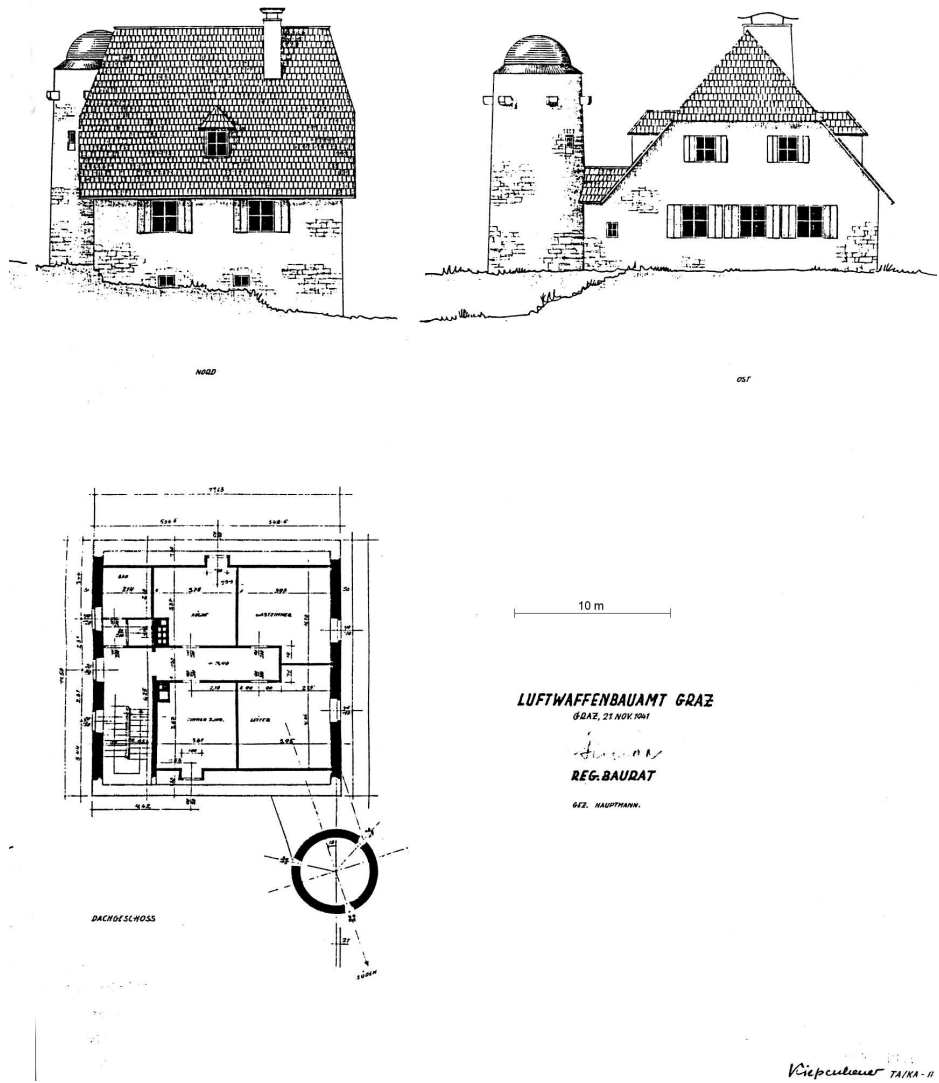}}
 \caption{Part of the original buildingplan from 1941 commissioned by the ``German
 Luftwaffenbauamt Graz" and signed by K.O.~Kiepenheuer (lower right corner).}
 \label{fig:plan}
\end{figure}

In the 1930s knowledge about radio wave propagation evolved, and 
it became obvious that the Earth's ionosphere is influenced by solar activity.
M\"ogel and Dellinger \citep{Dellinger1935} found that flares (``solar eruptions") can
increase the density of low-altitude layers of the ionosphere, which leads to
an absorption of short-wave radio waves causing blackouts of radio communications.
In WW\,II radio communication became an important and necessary means of communication
and navigation for airplanes and submarines as their operating distance was 
increasing \citep{Seiler2007}. Thus, the ``Deutsche Luftwaffe" (German Airforce) 
founded a network of observatories in the Alps, Zugspitze, Schauinsland, and Wendelstein 
in Germany and Kanzelh\"ohe (Figure \ref{fig:plan}) in Austria, 
to collect knowledge about these effects and to achive a better understanding of 
the solar-terrestrial relations. The goal was to inform the Luftwaffe in case of 
disturbances of the ionosphere or even to produce forecasts of such disturbances.\\
The location (N 46$^\circ$40.7$'$, 13$^\circ$54.1$'$, altitude 1526 m) of Kanzelh\"ohe
Observatory (KSO) was chosen because the area was reachable throughout the year by 
cable car, it had good observing conditions \citep{Eckel1937}, and it was located 
near a city (Villach). The scientific work was guided under the direction of K.~O. 
Kiepenheuer from the Fraunhofer Institute in Freiburg, Germany 
(now ``Kiepenheuer Institut", KIS). In autumn 1941 the construction works were 
started and in 1943 observations with state of the art equipment began. 
In addition the construction of a third observing dome for a more modern and 
larger coronagraph on the top of mount Gerlitzen began, but was not 
completed before the end of the war. More detailed information about the history 
of the solar research during the ``Third Reich" has been given by 
\cite{Seiler2007} and \cite{Kuiper1946}.\\
After WW\,II the observatory was reorganized and affiliated with Graz University as part
of a new institute. The official confirmatory was in the year 1949 after the founding of 
the second republic \citep{Jungmeier2014a,Jungmeier2014b}. 
The tower on the top of the Gerlitzen was occupied by the British Allied Forces,
and in return a new observation dome was constructed for solar corona observations near the
occupied tower. In 1965--66 the observatory building was reorganized and extended. 
The northern dome, which stood separately, was integrated to the main observatory building.
It was equipped with mechanical and precision engineering workshops and an optical 
laboratory. 
From 1989 to 1991 a further extension to the building was erected, which houses
the library, laboratories, and a fireproof and air-conditioned archive.\\
Solar observations at KSO started in 1943; the oldest data available from 
this time are white-light photographs of the Sun from July 1943. The first sunspot 
drawings date back to May 1944. In the beginning, during WW\,II the main communication 
and data transfer took place within the German network, but in 1948 
international cooperation began and data was transferred to Greenwich, Zurich, Freiburg,
Paris, Moscow, and other international data centres. \\
In this article we give a brief historical overview of the instruments and
observations at KSO since its foundation. Next attention is turned to the sunspot
observations and how the relative sunspot number at KSO is obtained. In this context
we lay the main focus on the comparison of the KSO sunspot numbers with the recently
recalibrated International Sunspot Number \citep[ISN:][]{Clette2014} and
discuss the reasons of discrepancies between these numbers for certain periods.

%%%%%%%  figure  figure  figure  figure  figure  figure  figure  figure %%%%%%%%

%%%%%%%%%%%%%%%%%%%%%%%%%%%%%%%%%%%%%%%%%%%%%%%%%%%
\section{Instruments and Observations: a Historic Overview}\label{s:instrumentation} 

\subsection{Instrumentation}

\begin{figure}
 \centerline{\includegraphics[width=1\textwidth,clip=]{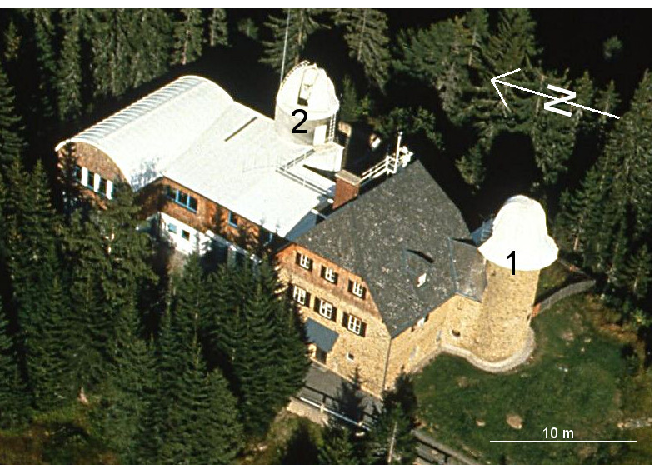}}
 \caption{Aerial view of the observatory around 1995. The original building
 is covered with a dark roof; the extensions were given a white roof to prevent air heating
 for better seeing conditions. Until 1973 the observations where performed in
 the southern tower (1). With the construction of the patrol instrument in the 
 northern tower (2) the observations in the southern tower were stopped.}
 \label{fig:aerial}
\end{figure}

There exist two eras of instrumentation at the observatory: the time before 1973,
where the main observations were performed in the southern tower (1 in Figure\ref{fig:aerial}) 
and the era of the patrol instrument in the northern tower (2).

%%%%%%%  figure  figure  figure  figure  figure  figure  figure  figure %%%%%%%%

\subsubsection{1943 -- 1973}

\paragraph*{Northern Tower}

Until 1947 a coronagraph \citep[{\it d/f} = 11/165\,cm,][]{Comper1957} was installed in 
the northern tower. This coronagraph was then brought onto the top of the Gerlitzen 
mountain (1911 m a.s.l.) into the new dome constructed by the British Allied forces. 
It was operated there until 1964 when the observing conditons became worse due to
a 25\,m radio tower newly erected near the dome. Also on this coronagraph a telescope 
for producing sunspot drawings with a diameter of 15\,cm was mounted piggyback.
Until 1958 a tiny 12\,cm refracting telescope with a camera was operated in this tower for
night observations, which on the occasion of the International Geophysical 
Year (IGY) in 1958 was equipped with an H$\alpha$ monochromator from Zeiss. From this time on
regular photographic H$\alpha$ observations have been made.

\paragraph*{Southern Tower}

\begin{figure} 
 \centerline{\includegraphics[width=0.6\textwidth,clip=]{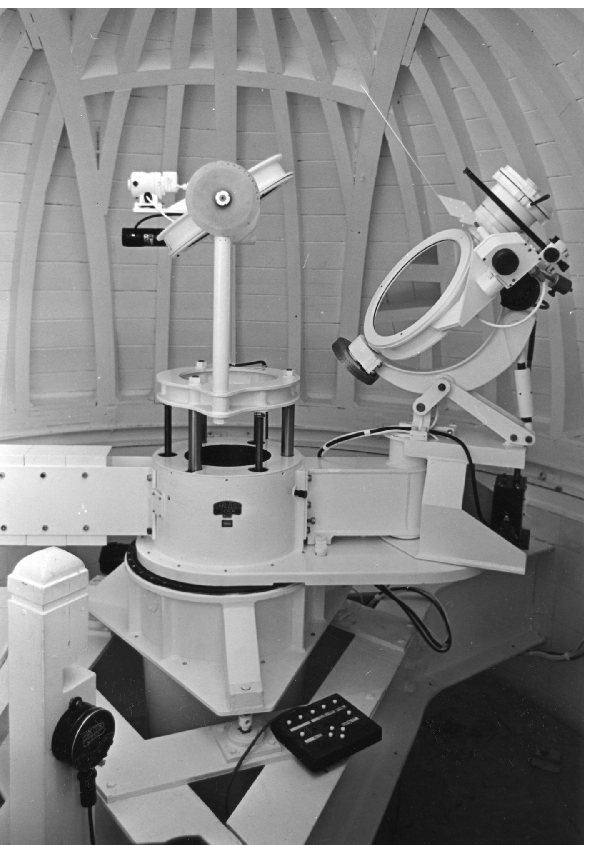}}
 \caption{The Zeiss-Heliostat with two flat mirrors of 30\,cm in diameter
 mounted in the southern tower was in use from 1943
 until 1973. The sunlight was reflected down to the laboratory in the basement, 
 where the drawing table and the spectrohelioscope were installed.}
 \label{fig:heliostat}
\end{figure}

A heliostat from Zeiss (see Figure \ref{fig:heliostat})
reflected the sunlight down to the laboratory (located in the basement) where all 
instruments were situated. The two flat mirrors of the heliostat had a diameter of 30\,cm and were
guided by a synchronous motor. The mirrors were originally silver-coated, which was 
not a long-lasting solution. Therefore in 1950 the mirrors were coated by an 
aluminium evaporation deposition, which was protected by a thin silica film.
In the 1960s the guiding of the heliostat was improved by installing a remote 
control and servomotors.

%%%%%%%  figure  figure  figure  figure  figure  figure  figure  figure %%%%%%%%

\begin{figure} 
 \centerline{\includegraphics[width=0.6\textwidth,clip=]{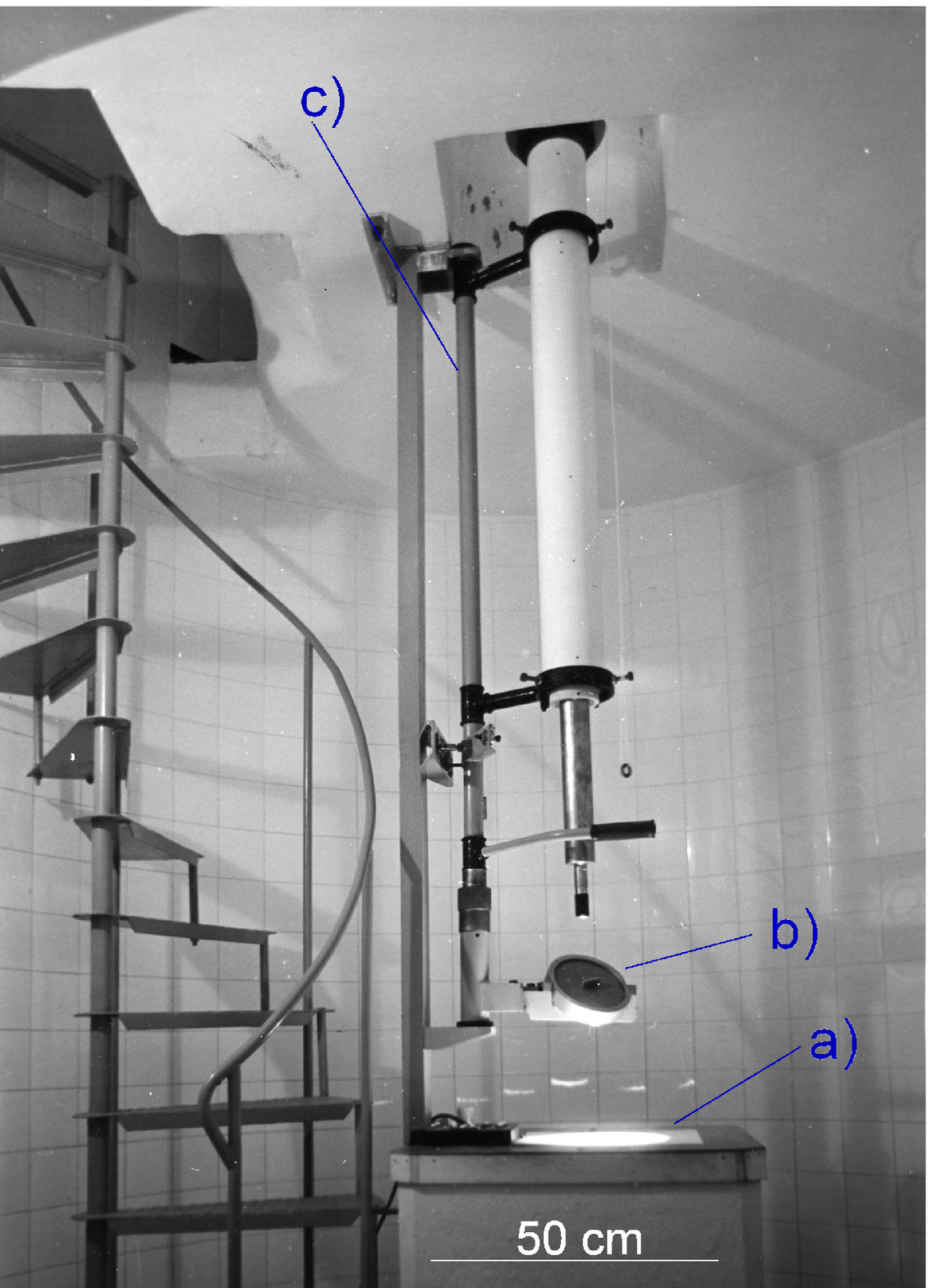}}
 \caption{The vertical telescope was mounted in the southern tower and produced
 a solar image with 25\,cm in diameter (a). The telescope could be pivoted
 araund a vertical axis (c) and the light beam was reflected by a 45$^{\circ}$
 inclined mirror (b) into the laboratory, where the spectrohelioscope was
 installed.}\label{fig:verttelescope}
\end{figure}

In the lower section of the tower a vertical telescope with an aperture of 11\,cm 
and a focal length of 165\,cm was mounted (Figure \ref{fig:verttelescope}). This
device produced a solar image of 25\,cm in diameter onto a drawing board fixed 
on a bricklaid pedestal. Until November 1946 the vertical telescope was mounted onto 
the coronagraph. By changing the ocular, the projected image size was enlarged 
to 25\,cm. The vertical telescope was pivot-mounted in order to move it out
of the light path and to collimate the sunlight via a 45$^{\circ}$ inclined mirror into
the laboratory.

%%%%%%%  figure  figure  figure  figure  figure  figure  figure  figure %%%%%%%%

\begin{figure} 
 \centerline{\includegraphics[width=1\textwidth,clip=]{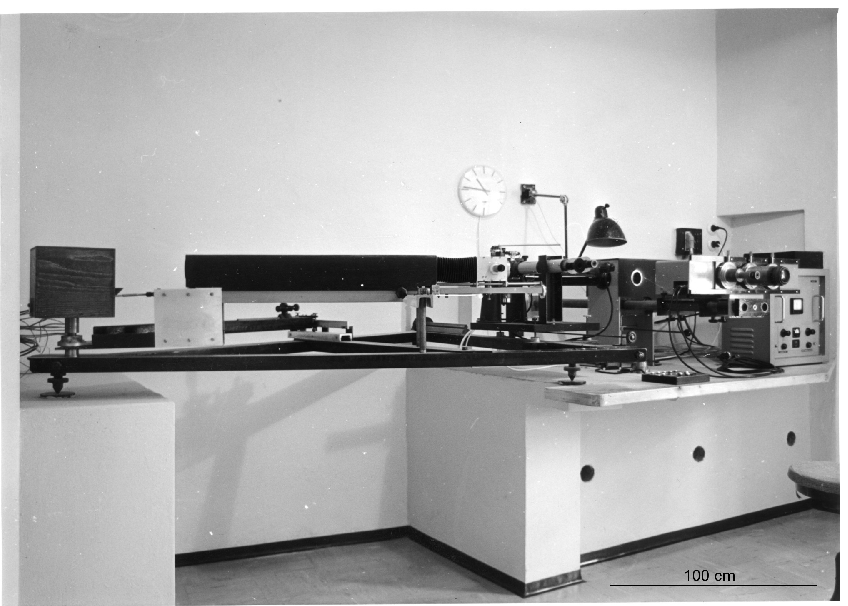}}
 \caption{In the basement laboratory a spectrohelioscope designed by Siedentopf
 was installed; here the Sun was observed visually in the chromospheric 
 H$\alpha$ line and the chromospheric features were added to the sunspot drawings.}
 \label{fig:spectrohelioscope}
\end{figure}

In the laboratory basement a spectrohelioscope (Figure \ref{fig:spectrohelioscope};
was operated. A schematic mode of operation of this device can be found 
in \cite{Siedentopf1940} and \cite{Comper1958}).
The observations at the spectrohelioscope were made visually and the chromospheric
phenomena observed in the H$\alpha$ line were added to the sunspot drawing.

%%%%%%%  figure  figure  figure  figure  figure  figure  figure  figure %%%%%%%%

\subsubsection{The Patrol Instrument: 1973 -- Today}

\begin{figure} 
 \centerline{\includegraphics[width=1\textwidth,clip=]{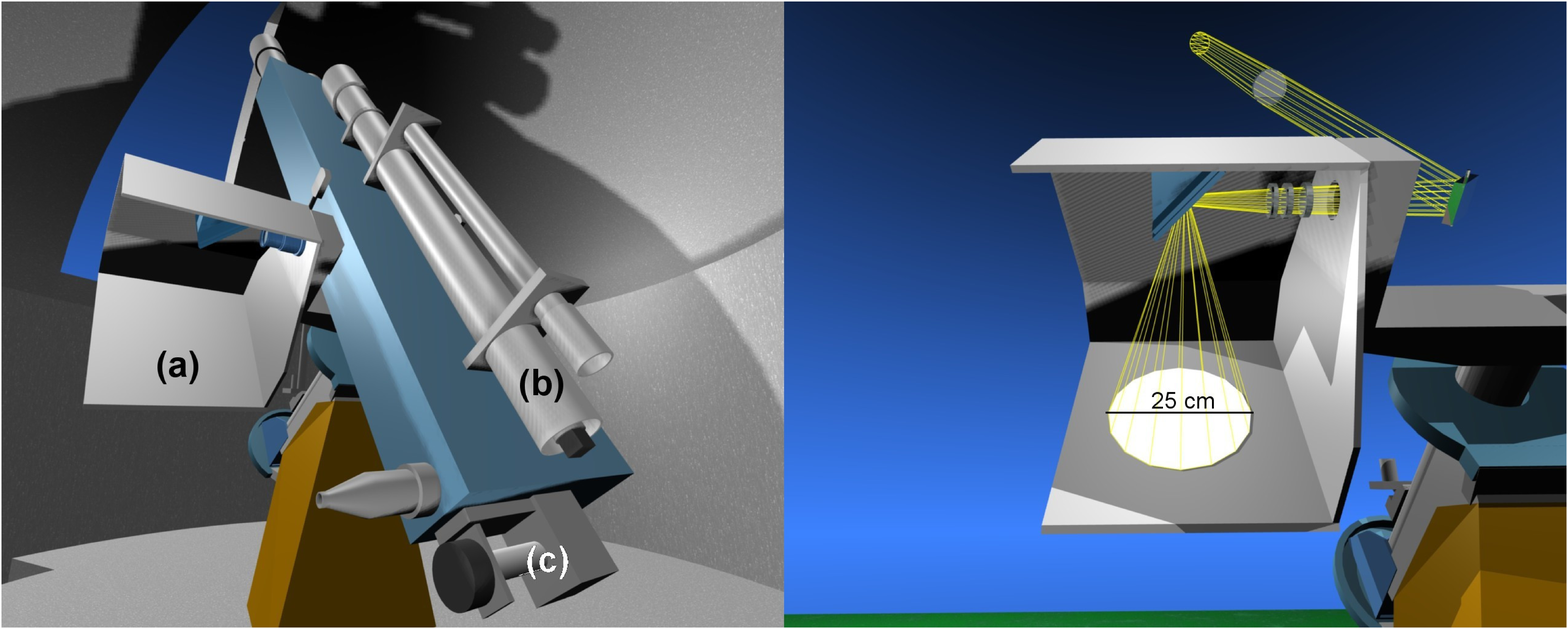}}
 \caption{Left: a schematic view of the patrol instrument comprising (a) the 
 drawing device, (b) the white-light telescope, and (c) the H$\alpha$ telescope.
 Due to the mirrors, the drawing is reversed left to right and because of the 
 light path through the declination axis of the telescope the image is also rotated by the 
 declination angle of the Sun.}
 \label{fig:drawdevice}
\end{figure}

The construction of the patrol instrument began in the mid 1960s during the
times when the observatory was enlarged and a number of interior works and
reconstructions were done, but it took until 1973 to complete the instrument 
and to shift the main observations there. The patrol instrument comprised 
three (later four) refractors on a common equatorial mounting. The diurnal movement 
was tracked by a microprocessor system, which was later on improved by a 
four-quadrant photocell controller and finally with the transfer to CCD cameras by 
using the solar disc image. The following instruments were mounted on the patrol instrument:\\
{\bf H$\alpha$ telescope:} in the beginning it was equipped with a miniature 
 film adapter that was controlled automatically so that every four minutes one 
 image was taken. The film rolls, each consisting of about 1000 images, were 
 completely digitized in 2007 \citep{Poetzi2007}. In 1998 the recording technology
 was changed to CCD cameras \citep{Hanslmeier2003}, which were upgraded in 2005 
 and 2010 \citep{Poetzi2015}.\\
{\bf Drawing device:} the objective lens of the old vertical telescope was 
  reused and a new zoom optics system was built in order to obtain 
  the same size of the projected solar disc as before (25\,cm). 
  A great benefit was the arrangement of 
  the drawing device directly on the declination axis of the telescope. Thus, the 
  observers position during drawing is the same as on a lectern 
  (see Figure \ref{fig:drawdevice}) and the forces applied by the observer have 
  minimal effects onto the telescope motion.\\
 {\bf White-light telescope:} beginning with 1989 \citep{Pettauer1990} images 
  were captured on on large size film (13\,cm $\times$ 18\,cm). The data and films are 
  currently being digitized \citep{Poetzi2010}. In 2007 a CCD camera replaced 
  the old system, which was again replaced in 2015 by a camera with more greylevels. \\
 {\bf Magneto Optical Filter:} this device \citep{Cacciani1999} was only 
  installed between 1999 and 2002, producing intensity images, dopplergrams, and
  magnetograms.\\
 {\bf Ca\,{\sc ii}\,K telescope:} This telescope was installed in 2010; the 
  filter is centered at 393.37\,\AA\, and it was operated from the beginning with 
  a CCD camera \citep{Polanec2011}.

Table \ref{tbl:instrum} gives an overview of the telescopes mounted on the patrol
instrument and the corresponding data products obtained over the years. Nowadays,
the following telescopes are still in use: Drawing device, H$\alpha$, white-light and
Ca\,{\sc ii}\,K; all are running in patrol mode observng the full-Sun with high
cadence. The sunspot drawing is made once 
a day, it is immediately scanned to be archived in the KSO archive system 
(\url{cesar.kso.ac.at}).

\begin{table}
  \caption{Overview of the telescopes mounted on the KSO patrol instrument and
  their data formats. Almost all photographic data is digitized, new drawings
  are scanned immediately.}
  \label{tbl:instrum}
  \begin{tabular}{lccccc}
    Instrument & \multicolumn{2}{c}{operating}  & Cadence & Size [pixel] &  \\ 
               & from & to &  & greylevels & \\ \hline{}
    Drawing device & 1973 &  & 1 per day &  &  \\
            & (1973 &  &  & 1700 $\times$ 1850 & digitized) \\
    H$\alpha$  & 1973  & 2000   &  240 sec       &  photographic &    \\
      & (1973  & 2000   &         &  1024 $\times$ 1024 / 256 &  digitized)  \\
      & 1998  & 2005   &  100 sec       &  1008 $\times$ 1016  / 256 &  \\
      & 2005  & 2010   &  10 sec       &  1024 $\times$ 1024  / 1024  & \\
      & 2010  &    &  6 sec       &  2048 $\times$ 2048 / 4096  & \\
    White light  & 1989  & 2007   &  3 per day       &  photographic &   \\
      & (1993 & 2007   &        &  2200 $\times$ 2200 / 32768 & digitized)  \\
      & 2007  & 2015   &  60 sec       &  2048 $\times$ 2048 / 1024  & \\
      & 2015  &    &  15 sec       &  2048 $\times$ 2048  / 4096  & \\
    MOF & 1999 & 2002 & 60 sec & 512 $\times$ 495 / 32768 & \\
    Ca\,{\sc ii}\,K & 2010 & & 6 sec & 2048 $\times$ 2048 / 4096 &  \\
  \end{tabular}
\end{table}

%%%%%%%  figure  figure  figure  figure  figure  figure  figure  figure %%%%%%%%

\subsection{The Sunspot Drawings}

\subsubsection{From May 1944 until November 1946}

\begin{figure}
 \centerline{\includegraphics[width=0.95\textwidth,clip=]{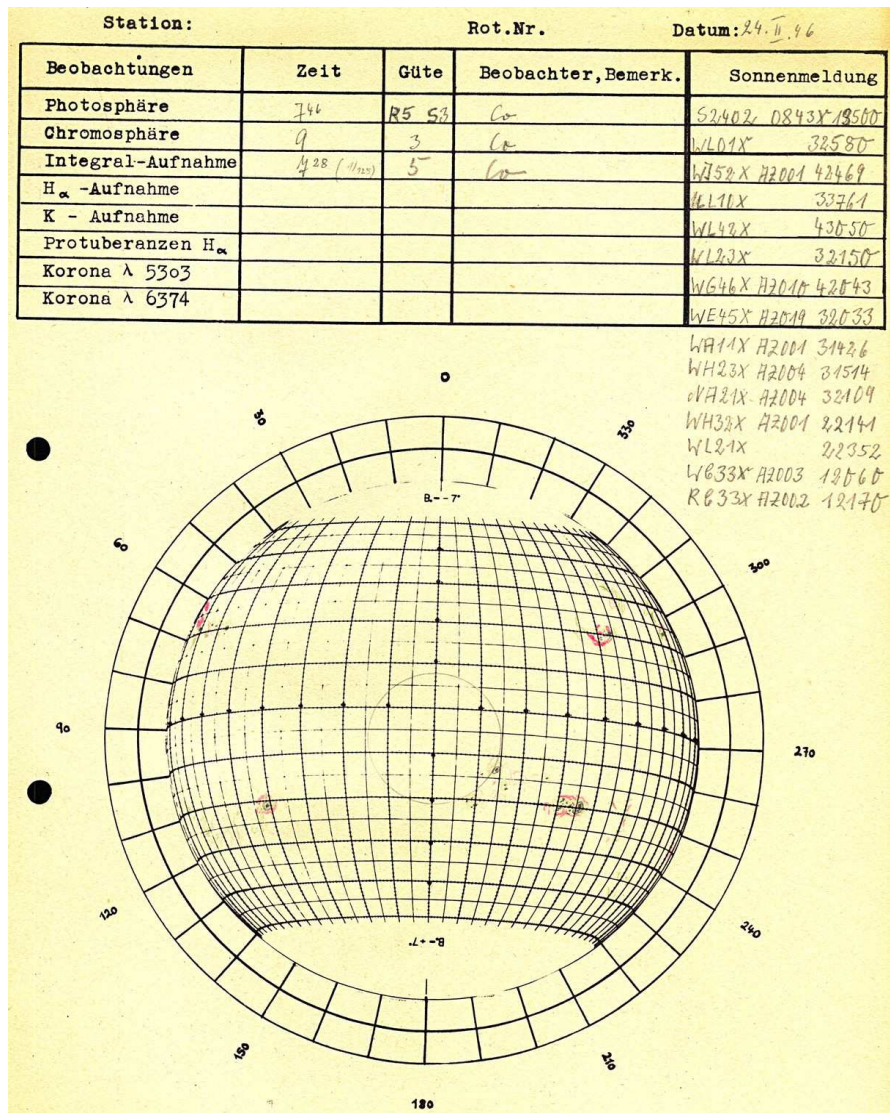}}
 \caption{Sample sunspot drawing from 24 February 1946. The German Airforce prepared
 heliographic coordinate sytem templates for the drawing available in one-degree
 steps for solar {\it B$_0$} angles. Chromospheric plages were added in red.}
 \label{fig:draw19460224}
\end{figure}

During WW\,II and in the first years after the war, the sunspot drawings were made 
in compliance with the German system, which was specified by K.O.~Kiepenheuer
(see Figure \ref{fig:draw19460224}).
The drawings were made with a piggyback telescope on the coronagraph
in the northern tower.
The solar disc on the templates was 15\,cm in diameter and the heliographic
coordinate system was pre-printed. There existed 15 different types of heliographic
coordinate templates, one for each degree of the latitude of the centre of the 
solar disc $B_0$. The grid for latitude and longitude was divided into five-degree steps. 
In addition to the sunspots, also plages observed in H$\alpha$ were also sketched in red.

%%%%%%%  figure  figure  figure  figure  figure  figure  figure  figure %%%%%%%%

\noindent In the header section of the sunspot drawings, the following information was given:
\begin{compactitem}
\item Type of observation 
\item Time of drawing (CET), of chromosphere observation (H$\alpha$), and
  of white-light photograph.
\item Quality for the observations, given as {\it R} (quietness) and {\it S} (sharpness) 
      between 1 (exceptional) and 5 (very poor), defined in an
      internal communication by \cite{Kiepenheuer1946}.
\item Name or initials of observer.
\item Position and code for daily solar report (``Sonnenmeldung").
\end{compactitem}
For a later check of the sunspot numbers, the ``Sonnenmeldung" is of great importance,
as it is impossible today to count the sunspots on the drawings directly. 
The Sonnenmeldung in Figure \ref{fig:draw19460224} has to be interpreted as listed 
in Table \ref{tbl:code}.

\begin{table}
  \caption{The code of the ``Sonnenmeldung" consisted of information on
  observing time, seeing conditions, sunspot groups, and faculae observed.
  Here we show the interpretation of some datablocks from Figure \ref{fig:draw19460224}.
  For each sunspot group or faculae the area of the faculae in the photosphere and
  chromosphere was also noted. Some identifiers could not be identified, like the first
  character in the first block, which can be W, P, R or T. }
  \label{tbl:code}
  \begin{tabular}{ccccccc}
     Line &\multicolumn{3}{c}{Datablocks}& \multicolumn{3}{c}{Interpretation} \\
     \hline   
     1 & 1 & 2 & 3 & date & time/quality & sun spot  \\
       & S2402 & 0843X & 19500 &  Feb. 24  & 8 MEZ             &  number\\
       &    &       &       &           & photosph. 4 &  {\it R} = 195 \\
       &    &       &       &           & chromosp. 3 &  \\ \hline
       & 1 & 2 & 3 & type and area & sunspots & position \\
     2  & WL01X & & 32580 & L=faculae  & & quadrant 3 \\
       &       & &       & photosph. 0 & & S25W80 \\
       &       & &       & chromosph. 1 & & \\ 
     3 & WJ52X & AZ001 & 42469 & J-group & 1 spot & quadrant 4  \\
      &       &       &       & photosph. 5  &        & N24W69 \\
      &       &       &       & chromosph. 2 & & \\ 
     8 & WE45X & AZ019 & 32033 & E-group & 19 spots & quadrant 3  \\
      &       &       &       & photosph. 4  &        & S20W33 \\
      &       &       &       & chromosph. 5 & & \\ 
     15 & RC33X & AZ002 & 12170 & C-group & 2 spots & quadrant 1  \\
      &       &       &       &  photosph. 3 &     & N21E70 \\
      &       &       &       & chromosph. 3 & & \\ 
  \end{tabular}
\end{table}

\subsubsection{From November 1946 until  May 1973}

\begin{figure}
 \centerline{\includegraphics[width=1\textwidth,clip=]{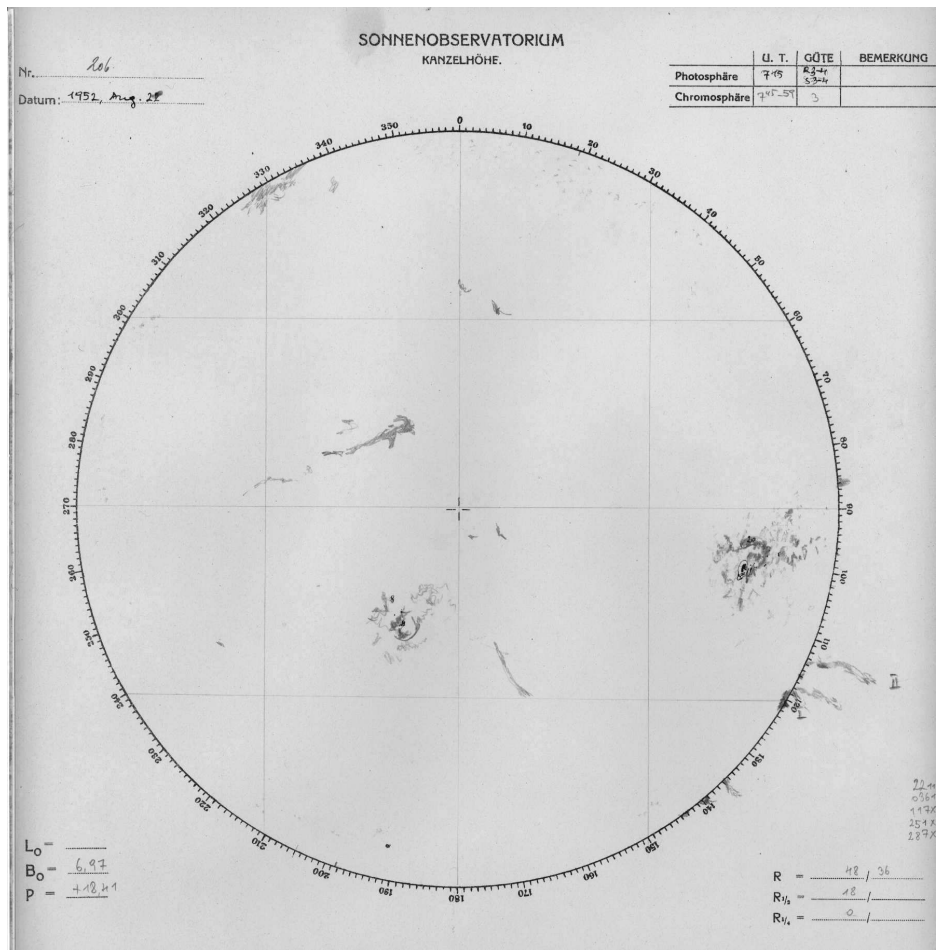}}
 \caption{
 Sunspot drawing from 22 August 1952: Besides the sunspots, also chromospheric 
 plages (around sunspots), photospheric faculae, and filaments were
 drawn. The chromospheric features were detected visually at the spectrohelioscope in
 the H$\alpha$ spectral line. In order to draw these features onto the correct place a 
 grid was added. Sometimes the locations and time of flares were also added.}
 \label{fig:19520822}
\end{figure}

In November 1946 a new projection lens was mounted onto the telescope in
order to obtain a projection of the solar disc with 25\,cm in diameter. 
This instrumental change from 15 to 25\,cm caused an increase in sunspot 
detection by $\approx$50\,\%. In November 1947 the telescope
was moved from the coronagraph to the southern tower, now called the ``vertical 
telescope", and the drawings were made on a stable pedestal. On the drawing 
templates, all sunspots, filaments, and faculae were drawn, sometimes even prominences. 
For the identification of chromospheric features the observer was looking through 
the spectrohelioscope and tried to draw these features as well as possible at the 
correct position of the sunspot drawing. For this purpose a rectangular grid was drawn onto 
the template (Figure \ref{fig:19520822}). The chromospheric plages 
were added in red, the photospheric faculae (only near the limb) in 
green, the filaments and visible lower parts of prominences were sketched in grey.\\
Until 1957, the sunspot numbers were also derived separately for the central zone, {\it i.e.} 
sunspots inside half of the solar radius. Until 1948 seeing conditions were not
taken into account for deriving the sunspot number. With the beginning of the year 
1952, the observation time was changed from CET to UT.

%%%%%%%  figure  figure  figure  figure  figure  figure  figure  figure %%%%%%%%

\subsubsection{From May 1973 To Date}

\begin{figure}
 \centerline{\includegraphics[width=1\textwidth,clip=]{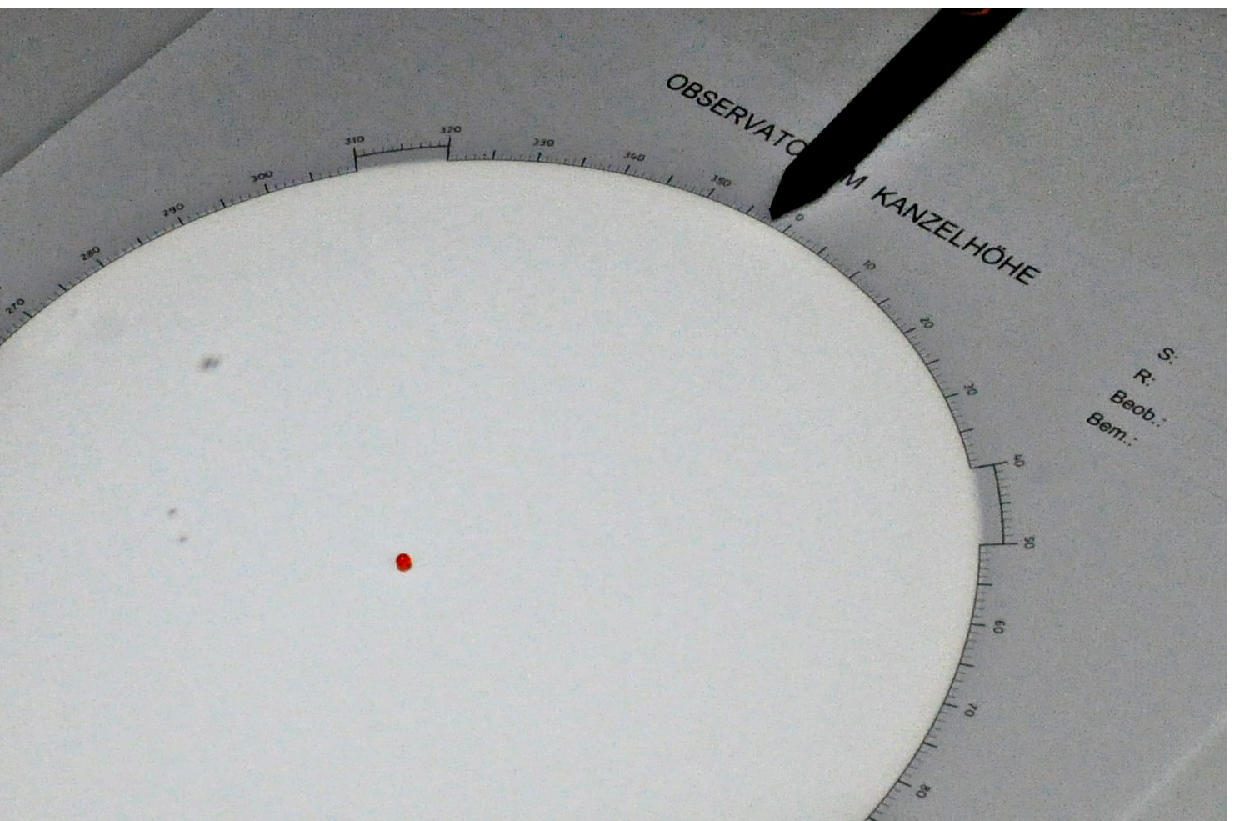}}
 \caption{A drawing template fixed in the drawing device. The red dot in the
 centre of the solar image is the needle, which is stuck into the template through
 the solar disc centre. The black arrow shaped holder in the upper part of
 the image fixes the template at the negative declination angle of the
 Sun.}
 \label{fig:drawing}
\end{figure}

From May 1973 to date the sunspot drawings have been made at the patrol instrument in the 
northern tower (Figure \ref{fig:drawing}). Chromospheric features are no longer  added to the drawings 
as in parallel the photographic patrol observations in H$\alpha$ began \citep{Poetzi2007}.

%%%%%%%  figure  figure  figure  figure  figure  figure  figure  figure %%%%%%%%

\subsection{International Cooperations}

Until 1945 the main cooperation took place with Freiburg and the
newly founded network of German observatories \citep{Seiler2007,Jungmeier2014a}. 
After WW\,II the southern part of Austria was occupied by the British allied forces.
A continuation of the cooperation with Freiburg was hardly possible as it was 
located in Germany. The local staff at KSO got into contact with the Astronomer 
Royal from the Greenwich Observatory,
which led to a fruitful cooperation and secured the existence of the observatory.
According to the \citeauthor{RAS1947} in \citeyear{RAS1947}, the cooperation 
with Greenwich had already begun in 1946.
The cooperation with other institutes began in 1948 according to the 
KSO activity reports \citep{KSO1946}. Sunspot numbers were sent 
to Freiburg each month, quarterly copies of sunspot drawings were
sent to Zurich, and flare data were sent to Meudon, France. From 1957 
on, the International Geophysical Year (IGY), data were sent to Freiburg every 14 days,
to Meudon, Pic du Midi, Boulder, and Moscow every month, and quarterly to Z\"urich
\citep{Mathias1962,Haupt1971}.
The activity reports contain no information about the cooperation with Greenwich,
which seems to have stopped with the end of the occupation of Austria by the 
Allied forces in 1955.\\
With new technologies (Telex in the 1970s and later internet) the data transfer 
to other institutes and data centres was improved and became more frequent. Nowadays 
the sunspot number is sent directly to the SILSO database (Sunspot Index and
Long-term Solar Observations in Belgium), and the H$\alpha$ patrol images are sent 
every night to the Global High Resolution H-alpha Network \citep{Steinegger2000} 
at the New Jersey Institute of Technology. Flare reports and patrol times are sent 
on a monthly basis to the National Centers for Environmental Information 
(NCEI\footnote{NCEI the world's largest active archive of environmental data was 
established in 2015 from the merger of the National Climatic 
Data Center (NCDC), the National Geophysical Data Center (NGDC), and the National 
Oceanographic Data Center (NODC).}), USA.
Long-exposure H$\alpha$ images showing off-limb prominence structures are 
sent to AISAS (Astronomical Institute of
the Slovak Academy of Sciences) as a supplement for the Lomnicky Stit prominence 
catalogue \citep{Rybak2011}. Web sites such as \url{solarmonitor.org}
display the Kanzelh\"ohe solar images. H$\alpha$ live images and real-time flare 
detections are provided via ESA's Space Weather Portal (\url{swe.ssa.esa.int}).
Additionally all observational data is also available via the online archive 
of the Kanzelh\"ohe Observatory \url{cesar.kso.ac.at} \citep{Poetzi2013}.

\section{Sunspot Numbers Derived at KSO}

The sunspot groups are classified according to the Zurich classification
scheme \citep{Waldmeier1955}, which describes the evolution of sunspot groups. 
The relative sunspot number is then obtained by counting the individual 
sunspots [{\it s}] and sunspot groups [{\it g}] as:
\[
  R = k(10\cdot g+s)
\]

The reduction factor [$k$] is a weighting factor that accounts in particular for the 
different telescopes, which is necessary when combining the data of different 
observatories. Until 2015 this 
factor was set to match the original 8\,cm telescope with a magnification of 64 
used by Rudolf Wolf in Z\"urich \citep{Waldmeier1961}.
{\bf  For the ISN a reduction factor
for each telescope is calculated, regardless of the observer and the observational 
conditions. But in principle for each observer an individual reduction factor
can be applied.}

\subsection{Reduction Factor $k$ at KSO}

\begin{figure}
 \centerline{\includegraphics[width=1\textwidth,clip=]{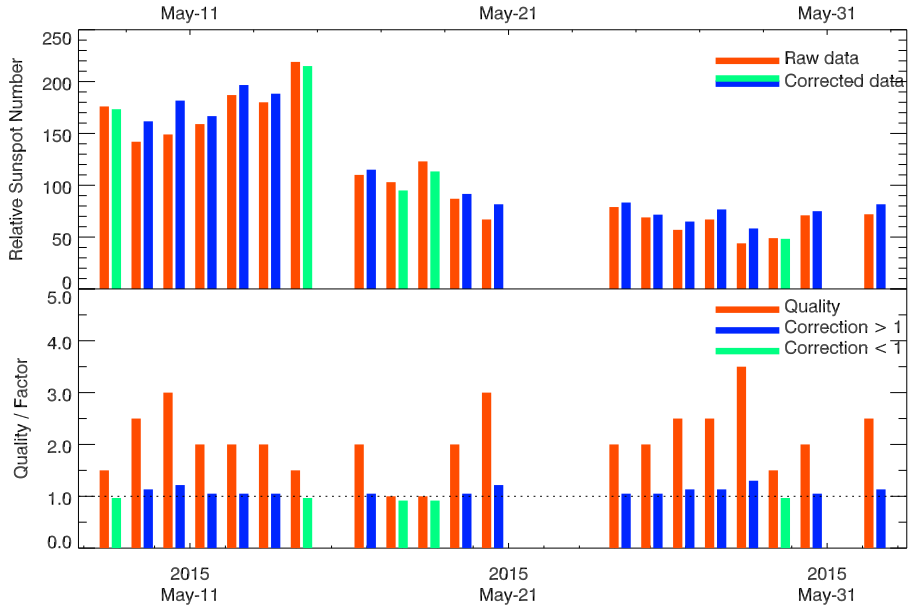}}
 \caption{Influence of the seeing conditions on the relative sunspot number derived. 
 In the lower panel the image quality for each day and the corresponding 
 correction factor is plotted. In the upper panel the raw relative sunspot 
 numbers (red) and the corrected sunspot numbers are plotted side by side.
 Blue (green) indicates relative sunspot numbers that are higher (lower) then the raw data.}
 \label{fig:corr}
\end{figure}

The atmosphere of the Earth influences the observation conditions. Air turbulence,
clouds, and wind have an impact on the quality; even a clear sky
is not a guarantee for good seeing. \cite{Haupt1965} compared seeing conditions
at KSO depending on general weather situations. He  showed,
{\it e.g.}, that the worst conditions occur when there is an upper air flow from the North 
and  a clear blue sky. Generally the best observing conditions 
are in the early morning, before the Sun heats up the ground, and also when there
are very thin high cloud layers.\\
The quality of the observations can be described by two main factors: the sharpness
and the quietness. At KSO the sharpness and the quietness have 
been used according to an internal communication by \cite{Kiepenheuer1946} and
redefined in \cite{Kiepenheuer1964}. The sharpness is defined as numbers between 
1 and 5 in steps of 0.5 depending on the details visible in the solar photosphere,
{\it e.g.}, 1 means that the granulation is clearly visible and even details inside the
umbra can be observed, whereas at a sharpness of 3 the granulation pattern is no longer 
recognizable.
The quietness, also between 1 and 5, describes the image motion inside the sunspots
and at the limb, {\it e.g.}, 1 stands for a completely stable image and at 3 the image motion 
is well visible on disc and at the limb but less than 4\,arcsec. The quality of the
observation is defined via the sharpness, as this parameter most strongly affects 
the number of visible sunspots. Low quietness makes the act of sunspot drawing 
more difficult as the projection is shaking. If the sharpness and the quietness become
worse, the number of sunspots that can be identified decreases, especially small A-spots
are no longer visible. On the other hand, the number of big H-spots is not affected
by the seeing. However, on average the number of observed sunspots increases with
better seeing quality. In order to obtain sunspot  numbers that are stable (in time)
and comparable (among different observatories) reduction 
factors for the seeing conditions have been introduced. Figure \ref{fig:corr} shows
how these reduction factors influence the raw sunspot number (red): 
If the quality (sharpness) is below 2 the corrected relative sunspot number 
(green) is smaller, whereas in the other cases the corrected relative sunspot number is larger than the 
raw relative number (blue). In general the dispersion of the daily relative sunspot 
numbers is reduced by the application of the reduction factor [$k$].

%%%%%%%  figure  figure  figure  figure  figure  figure  figure  figure %%%%%%%%

In the first years until 1948 the relative sunspot numbers obtained from the sunspot drawings 
have not been reduced. In 1948, Anton Bruzek (head of the observatory and observer 
during the years 1947 to 1953) made a first 
attempt to derive the reduction factors for the local observations \citep{KSO1946}. 
For this purpose he analysed all relative sunspot numbers obtained at KSO from 
1946 to 1948 and classified 
them according to the quality into classes from 1 to 5. He assumed that the relative
sunspot numbers in each class should be homogeneously distributed, i.e. the mean 
sunspot number in each class should be the same, if there was no influence by seeing 
conditions. With this method 
he calculated the correction factors between the different classes. In a first step he 
set the reduction factor for quality 3 to 1.0. In order to connect the local 
observations at KSO to the international sunspot numbers he used the observations from 
Z\"urich, Freiburg and the American Relative Sunspot Numbers [$R_{a}$: \citealp{Shapely1949}]. 
As the number of observations in the classes 1 and 5 was
very low, these factors are the most uncertain ones. Table \ref{tbl:factor}
lists the reduction factors used at KSO since 1946. In the first two years (1944\,--\,45)
no reduction factors were applied. Due to instrumental changes the factors had to be
recalculated twice, the first time in 1958 \citep{Haupt1959} and then again in 1979
\citep{Schroll1979}. Both times the recalculation was delayed by some years as 
there was a solar minimum and thus not enough days available with high relative 
sunspot numbers to obtain good statistics.
{\bf In 2015 with the recalibration of the relative sunspot numbers by \cite{Clette2014}
the ISN was adjusted to modern telescopes, {\it i.e.} the ISN was not only corrected for
{\it e.g.}, the transition form Wolf to Wolfer (factor 1.67) or Z\"urich weighting after 1947 (-18\%),
but also the whole series was divided by 0.6.}
 
\begin{table}
  \caption{KSO reduction factors [$k$] for the observed sunspot numbers according to the 
  seeing conditions. The initial $k$ factors were calculated in 1948, and were
  updated twice (in 1958, 1979) to account for instrumental changes. The update in 2015
  (multiplication of $k$ by $\frac{5}{3}$) is due to the recalibration of the ISN
  by \cite{Clette2014}. In the table we list the factors, the total numbers of
  sunspot drawings for each quality class and for the period of 1946 to 1948 the 
  mean sunspot number of each quality class.}
  \label{tbl:factor}
  \begin{tabular}{lccccccccc}
     Quality & 1 & 1\,--\,2 &2 & 2\,--\,3 &3 & 3\,--\,4 &4 &4\,--\,5 & 5 \\
     \hline   
     total (1946\,--\,1948)& 3 & & 44 & & 134 & & 121 & & 35 \\
     $\overline{R}$ & 390 & & 288 & & 222 & & 180 & & 150 \\
     $k$ (1948) & 0.38 & 0.46 & 0.52 & 0.59 & 0.67 & 0.75 & 0.86  & 0.97 & 1.00 \\ \hline
     total (1955\,--\,1957)& & 2 & 19 & 57 & 181 & 232  & 205 & 73 & 85 \\
     $k$ (1958) & 0 & 0.55 & 0.63 & 0.71 & 0.79 & 0.87 & 0.95 & 1.02& 1.10 \\ \hline
     total (1973\,--\,1978)& & 100 & 368 & 272& 212 & 146  & 77 & 32 & 34 \\
     $k$ (1979) & 0.55 & 0.59 & 0.63 & 0.68 & 0.73 & 0.79 & 0.85 & 0.92 & 0.99 \\ \hline
     $k$ (2015) & 0.92 & 0.98 & 1.05 & 1.13 & 1.22 & 1.32 & 1.42 & 1.53 & 1.65 \\ \hline
   \end{tabular}
\end{table}

\begin{figure}
 \centerline{\includegraphics[width=1\textwidth,clip=]{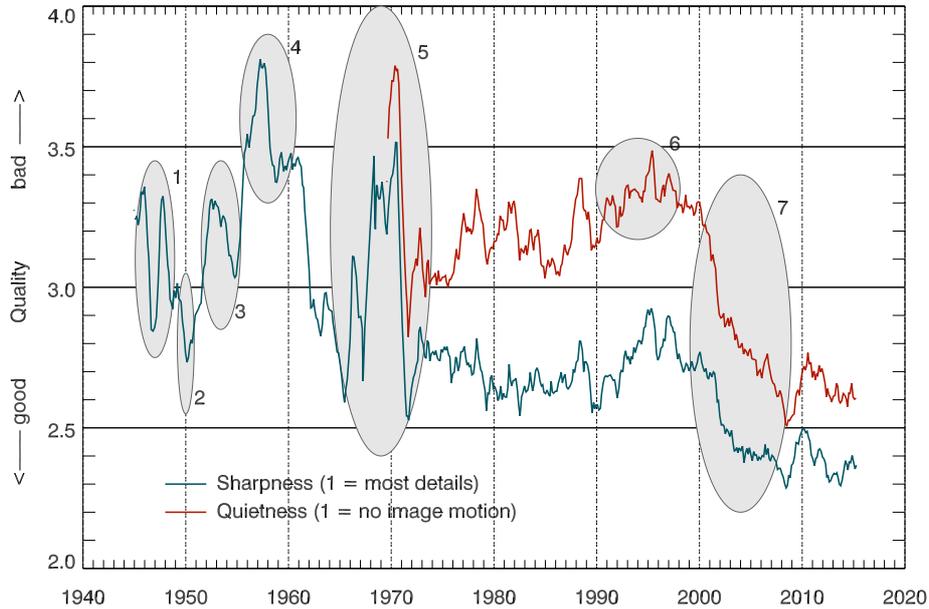}}
 \caption{Yearly running mean of sharpness and quietness values of the last 7 decades,
 the quietness is almost proportional to the sharpness.
 The ellipses mark times, when major changes in the observational conditions occurred.}
 \label{fig:seeing}
\end{figure}

Figure \ref{fig:seeing} plots the yearly running mean image quality from 1945 to 2015,
illustrating the change of the seeing conditions at KSO over
the past seven decades of observations.
After 25 years of quite unstable conditions until around 1970 they became more
stable for almost 30 years. From 2000 on, both the sharpness and the quietness of the images improved by 
at least half of a class. In Figure \ref{fig:seeing} rapid changes in the quality are marked with
ellipses and numbers and are discussed below:
\begin{itemize}
   \item[1] Due to the increase of the size of the projected image from 15\,cm to 25\,cm 
      in Nov. 1946, a larger number of sunspots could be identified in the
      observations. As the projection changed from the piggyback 
      instrument on the coronagraph to the projection in the southern tower in the
      year 1948 with a stable drawing table the quality of the observation further improved.
   \item[2] According to the activity reports there were problems with the guiding 
      of the heliostat in 1948 and problems with the aluminium coating of its mirrors.
      In the year 1950 a completely new coating of the mirrors enhanced the brightness and 
      therefore the contrast and quality of the projection. 
   \item[3] The quality of the mirrors degraded and due to the minimum of the
      solar cycle there were only a few sunspots visible, which makes a good estimation
      of the image quality difficult.
   \item[4] For the low quality values in 1957\,--\,58 no instrumental cause could be
      found in the activity reports. During this period the KSO observers detected a strong
      deviation of the local sunspot numbers from Zurich and Freiburg, and therefore
      they recalculated the reduction factors.
   \item[5] Time of major reconstruction works at the observatory and a period of
      frequent replacement of personnel. This can be seen in Figure \ref{fig:newobs},
      where we plot for each year the total number of observers as well as the 
      newly instructed observers. In the year 1965 many trees around the observatory 
      were cut,  especially in the principal wind direction, which may have led to better seeing conditions.
      Some modifications at the end of the year 1966 in the southern tower led 
      to improved observation conditions as the airflow changed. 
   \item[6] For a few years the seeing conditions became worse as a result of the 
      Pinatubo volcano eruption in June 1991 \citep{Otruba1993}.
   \item[7] This increase in quality cannot be explained by improvements of the
      instrumentation or by any changes in the vicinity of the observatory. We may 
      speculate that climatic changes led to different atmospheric influences, but according to
      \cite{Auer2007} massive changes in temperature and air humidity in southern
      Austria had already begun in 1970.
\end{itemize}

%%%%%%%  figure  figure  figure  figure  figure  figure  figure  figure %%%%%%%%

\subsection{Comparison of the Kanzelh\"ohe Relative Number to the ISN}

%%%%%%%  figure  figure  figure  figure  figure  figure  figure  figure %%%%%%%%
\begin{figure}
 \centerline{\includegraphics[width=1\textwidth,clip=]{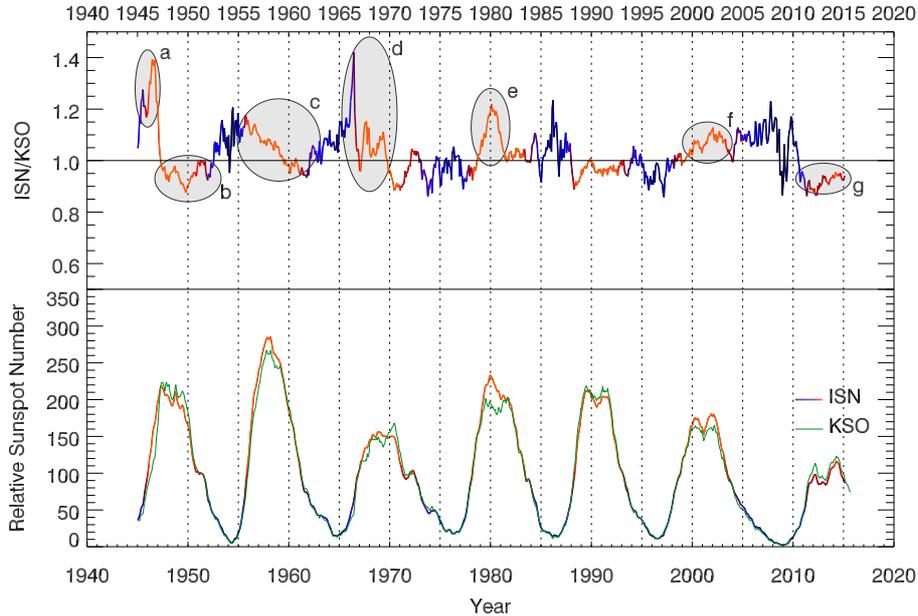}}
 \caption{Top: Ratio of the 13-month running mean of the recalibrated ISN \citep{SIDC}
 to KSO sunspot numbers. Values greater than unity indicate that the ISN is higher than
 the KSO sunspot number.  The color changes from blue to
 red for relative sunspot numbers above 100, {\it i.e.} solar activity minima are blue.
 Bottom: Relative sunspot numbers from KSO (green) and ISN (blue\,--\,red) 
 from 1945 to 2015.}
 \label{fig:kso_sidc}
\end{figure}

\begin{figure}
 \centerline{\includegraphics[width=1\textwidth,clip=]{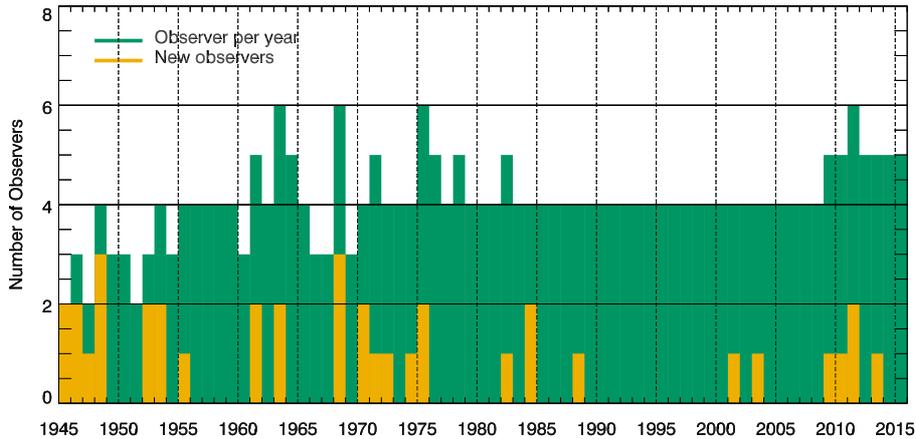}}
 \caption{The total number of observers for each year (green) and the number of 
  newly employed observers (yellow). On average, four observers are needed for 
  continuous observations, {\it i.e.} also on weekends and holidays. In former times
  fewer observers were necessary, as the weekly working time was 48 hours until 1958
  and then step by step reduced to 40 hours in 1979.}
 \label{fig:newobs}
\end{figure}

\begin{figure}
 \centerline{\includegraphics[width=1\textwidth,clip=]{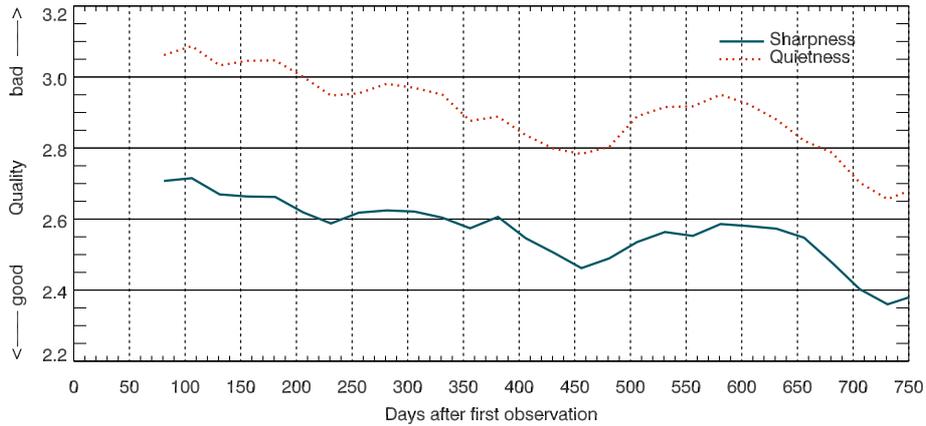}}
 \caption{Mean sharpness and quietness estimates derived from all individual 
 observers who contributed with more than 100 drawings (total number 21) during their first two 
 years of observing duty. The data is smoothed over half a year. Both values
 are higher in the first few months. An underestimate of the image quality causes 
 an overestimate of the sunspot numbers.}
 \label{fig:seeing_start}
\end{figure}

Figure \ref{fig:kso_sidc} shows the 13-month running mean sunspot numbers derived
at the KSO together with the International Sunspot Numbers 
(ISN; Silso World data Center, \cite{SIDC}) for the period 1945 to 2015.
The top panel shows the ratio of the two time series. In general both time series 
agree within a limit of $\approx$20\,\% (with two exceptions). The mean of the ratio of the ISN to 
the KSO sunspot numbers is 1.025 $\pm$ 0.088, {\it i.e.} the rms differencees are 
at a level of 9\,\%. In Figure \ref{fig:kso_sidc}, we also
indicate periods of larger deviations. However we do not consider relative differences 
at solar activity minima (blue), as the small sunspot numbers may lead to relatively
large deviations in the relative differences although the absolute agreement is 
very high. We note the following periods of larger ($\gtrsim$10\,\%) deviations:
\begin{itemize}
   \item[a,b] The same reasons as above in items 1 and 2 apply here. Additionally
    Anton Bruzek introduced a new reduction factor.
   \item[c] The drift between 1956 and 1962 cannot be explained by any instrumental
   changes. In 1958 new correction factors were calculated as it became clear
   that there was some deviation; these factors were higher and may be the reason 
   for the extension of the drift until 1962.
   \item[d] Between 1965 and 1968 major construction works were carried out at the 
   observatory, the observations were even stopped for some months in 1966, 1967,
   and 1968. A fluctuation of observers began in 1968, when
   three new observers were employed. These fluctuations lasted until 1975 with a total of 
   ten new observers (Figure \ref{fig:newobs}).
   \item[e] New reduction factors were used from June 1979 on, which led to 
   smaller sunspot numbers. The new $k$-factors for qualities 3 to 5 were about 10\,\% 
   lower than the old factors ({\it cf.} Tab. \ref{tbl:factor}); therefore, as the mean 
   quality was above 3, these factors should be responsible for a reduction of only 10\,\%.{}
   {\bf This deviation maybe due to the fact that there is also some uncertainty in
   the ISN series during this time as there was the closing of the Zurich observatory
   and the transition to Locarno as reference \citep{Clette2014}.}
   \item[f]  New observers were employed but also sudden image quality changes
   happened (see Figure \ref{fig:seeing}). Two observers went into retirement; their
   eyesight may have become worse causing some drift in the sunspot number,
   similar to the reason for the Locarno drift discussed in \cite{Clette2014}.
   \item[g] From 2009 new observers came to the observatory and were
   introduced into the observational work. New observers tend to underestimate the 
   image quality which results in higher sunspot numbers. Figure \ref{fig:seeing_start}
   shows how on average the image quality is estimated over the first two years 
   by new observers; the estimation is nearly half a class worse at the beginning
   of their career as observer.
\end{itemize}

%%%%%%%  figure  figure  figure  figure  figure  figure  figure  figure %%%%%%%%

%%%%%%%  figure  figure  figure  figure  figure  figure  figure  figure %%%%%%%%

\section{Discussion}

In general, the sunspot numbers derived at KSO and the ISN reveal a good agreement.
The mean of the ratio of the ISN/KSO monthly mean sunspot numbers for 
1945\,--\,2015 is 1.025 $\pm$ 0.088. However, there are a few periods where the relative 
difference exceeds $\pm$ 20\,\% (28 months out of 842). The main reasons for these 
large differences are instrumental improvements and observer fluctuations.
For some periods with major deviations ({\it e.g.} around 1980) no explanation could be found. 

\begin{figure}[H]
 \centerline{\includegraphics[width=1\textwidth,clip=]{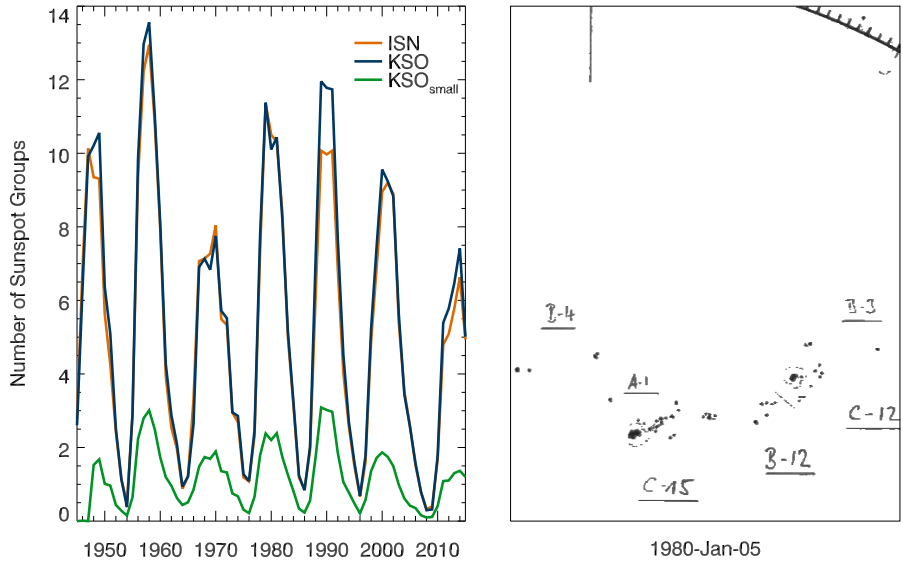}}
 \caption{
 Left: Comparison of yearly group numbers between the international 
 (red, \cite{Svalgaard2015}) and KSO (dotted blue) group numbers. In dashed green we plot the yearly average 
 number of small groups, {\it i.e.} A-1 to A-3, B-2 and B-3.
 In the right panel there is a display detail of a drawing showing the splitting of groups.}
 \label{fig:groups}
\end{figure}

An important factor in determining the  Sunspot Number 
is the group number \citep{Hoyt1998a,Hoyt1998b}, as each group counts as much as 
ten individual sunspots. A few years after the reduction factors of the new patrol 
instrument were recalculated by \cite{Schroll1979} he found out that there was still 
a large deviation from the ISN. His assumption was that the
number of groups could be too low as a result of assigning too many sunspots to
one group or the insufficient detection of small groups. 
Fig \ref{fig:kso_sidc} shows that around 1990 the KSO sunspot numbers agree well
with the ISN, but inspection of Figure \ref{fig:groups} shows that especially in 
this time the number of detected sunspot groups was considerably higher at KSO.
In 1980, when the KSO sunspot numbers were too low by $\sim$20\%, the number of 
sunspot groups detected at KSO was identical to the ISN sunspot group number.
The right panel in Figure \ref{fig:groups} shows an extreme
example of group splitting; another observer could have found only three groups 
in the same drawing, as it is not always clear how to split these groups.
It was drawn at a time when a video system was installed at the observatory that
displayed a live H$\alpha$ image, which made it easier to find the individual active
regions. In the last years this has become much easier, spaceborne
instruments such as the Solar and {\it Heliospheric Observatory} (SOHO) or the 
{\it Solar Dynamics Observatory} (SDO) provide the observers with
high resolution images in various wavelengths and additionally with magnetic maps.
Using such additional data is actually also a change in the method and could lead to
differences in the sunspot numbers derived that are no longer comparable to the long 
time sunspot data series anymore.

%%%%%%%  figure  figure  figure  figure  figure  figure  figure  figure %%%%%%%%

When new observers are trained, we have noticed that they try to do their best, and they
often tend to find more small sunspots than there are actually on the solar disc.
In cases of very good seeing conditions, they are not really able to distinguish 
between small sunspots and pores or big dark intergranular regions. On the 
other hand, they also tend to overlook new spots close to the limb. In the KSO database 
there is an extra entry for small sunspot groups, {\it i.e.} sunspot groups of the 
classes A-1 to A-3, B-2 and B-3. This number is also plotted in Figure \ref{fig:groups} (green) in 
addition to the total group numbers. This plot shows that the number of small sunspots 
is proportional to the total group number. Before 1947 small sunspot groups are 
missing due to the projection size of only 15\,cm and around 1990 a large number of small 
sunspot groups where detected at KSO. However, especially around 1990 there was no change 
in the observing team.

Starting with the IGY (1958) until 1964, each observatory got a certain time slot 
for intensified observations. The observers were complaining that they had to 
observe around Noon, which was very late as the seeing conditions are the best 
early in the morning. As the KSO operates a meteorological station for the
Central Institute for Meteorology and Geodynamics  (ZAMG) hourly 
sunshine information is available for all years. These sunshine data show that on 
average the Sun shines two hours before the drawing is made. This value did not 
increase during the IGY, so only the chromospheric observations were shifted to 
the determined time slot. The image quality even improved during this time 
(Figure \ref{fig:seeing}).
Only between 1966 and 1970 was the sunshine duration more than three hours before 
drawing the sunspots, which was probably a result of the personnel situation 
in this period. For three months in 1969 there was only one person at the 
observatory. The relatively bad seeing conditions reported during this period may 
thus be related to the fact that the sunspot observations were carried out 
one hour later than in general. {\bf In additon to the later observation time
the fluctuation of observers, the construction works, and the cutting of trees 
in the surrounding of the observatory may also have affected the quality 
estimation.}

Another impact could come from climatic changes; in the southern part of Austria
the air became dryer and the temperature rose by about one degree over the last
40 years \citep{Auer2007} and according to the studies of \cite{Haupt1965} 
also the general weather situation influences the image quality. Figure 54 of \cite{Clette2014}
shows the variation of the annual average quality index of Locarno. There the construction
of a building near the observatory led to a quality jump but there appears to be a 
slow change to lower quality values over the last 45 years.

We conclude that the relative sunspot number derived at KSO is in good agreement
with the ISN; there is no long-term drift between the two numbers. However, there are 
also periods with larger deviations that can be explained by new personnel,
instrumental changes, or modifications in the vicinity, and there are also periods
of deviations for which no reason was found. As the $k$ factors also depend on
the observers themselves, they should be recalculated and verified at regular
intervals, especially when an observing team does not change for a longer
period of time.

\section*{Disclosure of Potential Conflicts of Interest}

The authors declare that they have no conflicts of interest.

%%%%%%%%%%%%%%%%%%%%%%%%%%%%%%%%%%%%%%%%%%%%%%%%%%%%%%%%%%%%%%%%%%%%%%%%%%%
%% Acknowledgements
%
 \begin{acks}
    Many pieces of information about historical and instrumental details were not
    available in printed or written form. We are grateful to  Hermann Haupt and 
    Thomas Pettauer, who have shaped the observatory through their work
    since the early 1950s and 1960s, respectively, for all of the information that they provided 
    us through private communications that would have been lost otherwise.
 \end{acks}

%%% %%%%%%%%%%%%%%%%%%%%%%%%%%%%%%%%%%%%%%%%%%%%%%%%%%%%%%%%%%%
%% Bibliography
%
% Using BibTeX
%
 \bibliographystyle{spr-mp-sola}
 \bibliography{WP_2015_Bib}  
\end{article} 
\end{document}